\documentclass[showpacs,pra,amsmath,amssymb,aps,superscriptaddress]{revtex4}
\usepackage[dvips]{graphicx}

\newcommand{\bra}[1]{\langle#1|}
\newcommand{\ket}[1]{|#1\rangle}

\newcommand{\expect}[1]{\langle#1\rangle}

\newcommand{\var}[1]{\sigma^2(#1)}

\newcommand{\eq}[1]{Eq. (\ref{#1})}
\newcommand{\fig}[1]{Fig. \ref{#1}}
\newcommand{\sect}[1]{Sect. \ref{#1}}

\def\d{\operatorname{d}}\def\<{\langle}\def\>{\rangle}
\def\Tr{\operatorname{Tr}}

\def\arg{\operatorname{arg}}
\def\bit{\operatorname{bit}}

\begin{document}
\title{Purification of noisy quantum measurements}
	\author{Michele \surname{Dall'Arno}}
	\author{Giacomo Mauro \surname{D'Ariano}}
	\affiliation{Quit group, Dipartimento di Fisica ``A. Volta'', via Bassi 6, 27100 Pavia, Italy}
	\affiliation{Istituto Nazionale di Fisica Nucleare, Gruppo IV, via Bassi 6, 27100 Pavia, Italy}

	\author{Massimiliano F. \surname{Sacchi}}
	\affiliation{Quit group, Dipartimento di Fisica ``A. Volta'', via Bassi 6, 27100 Pavia, Italy}
	\affiliation{Istituto di Fotonica e Nanonetcnologie (IFN-CNR),
          P.le Leonardo da Vinci 32, I-20133 Milano, Italy}
	
	\date{\today}
	
	\begin{abstract}
          We consider the problem of improving noisy quantum
          measurements by suitable preprocessing strategies making many noisy detectors equivalent to a single ideal detector. For
          observables pertaining to finite-dimensional systems (e.g.
          qubits or spins) we consider preprocessing strategies that
          are reminiscent of quantum error correction procedures and
          allows one to perfectly measure an observable on a single
          quantum system for increasing number of inefficient
          detectors. For measurements of observables with unbounded
          spectrum (e.g. photon number, homodyne and heterodyne
          detection), the purification of noisy quantum measurements can be achieved by preamplification as suggested by H.
          P. Yuen \cite{yuen}.
	\end{abstract}
	
	\maketitle
	
	\section{Introduction}
	
	In many situations it is necessary to measure an
        observable in the presence of noise, e.g. when transmitting a quantum state
        through a noisy quantum channel that degrades it exponentially
        versus distance, corresponding to a degradation of the measurement.

        A number of figures of merit can be used to characterize the
        noise of non-ideal measurements.  An example of such figures
        of merit is the variance of the outcomes distribution.  An
        extensive analysis of the variance affecting quantum
        measurements has been done for example in \cite{ozawa}.  In a
        communication scenario, a relevant figure of merit is
        represented by the mutual information between the measurement
        outcomes and the input alphabet encoded on an ensemble of
        states.  The problem of how much classical information can be
        extracted from a quantum system has been first deeply
        discussed by Holevo \cite{holevo}, who provided bounds on the
        accessible information, and then revisited in the framework of
        quantum information by Schumacher et al.\cite{schumacher}.  A
        further figure of merit is the average probability of
        correctly distinguishing input states picked up from a given
        ensemble.  This is one of the first problems faced by
        quantum estimation theory, and has been addressed extensively
        in the literature \cite{holevo, hel, fuchs, buscemi06}.
        Finally, another example of figure of merit is a suitable
        distance between the noisy and the ideal outcomes probability
        for fixed input states.

\begin{figure}[htb]
		\includegraphics{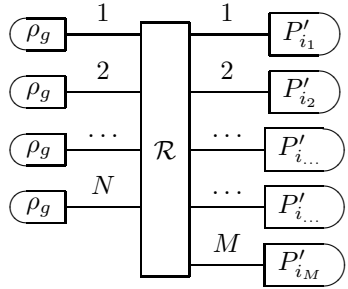}
		\caption{Purification scheme for noisy quantum measurements.}
		\label{fig:purification_scheme}
	\end{figure}

	In this paper, we consider the situation where $N$ identical
        preparations of the state $\rho_g$ belonging to some ensemble
        $\mathcal{S} = \{(p_g,\rho_g)\}$ are given. We are allowed to
        use $M$ non-ideal detectors, with $M \geq N$. Each detector is
        described by a Probability Operator-Valued Measure (POVM),
        namely a set of positive operators $\{P'_i\}$, which provides
        a resolution of the identity, i.e. $\sum _i P'_i =I$. Each POVM
        element $P'_i$ is the noisy version of an ideal POVM element
        $P_i$.  A generic quantum channel $\mathcal{R}$ is allowed to
        act on the $N$ identical copies of the state $\rho_g$ before
        the $M$ noisy POVM $\{P'_i\}$ are measured, and a generic
        classical post-processing can be done on the outcomes of such
        measurements.  Such a scheme of ``purification'' of noisy measurements is
        depicted in \fig{fig:purification_scheme}.
	We address the problem of optimizing the quantum channel $\mathcal{R}$ in order to reduce the effect of noise affecting the POVMs $\{P'_i\}$. 
	We approach the problem through the minimization of the variance of the maximum likelihood estimator for the parameter $g$ and through the maximization of the mutual information between $g$ and the measurement outcomes.
	
\begin{figure}[htb]
		\includegraphics{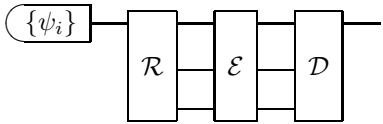}
		\caption{Scheme for quantum error correction.}
		\label{fig:error_correction_scheme}
	\end{figure}

	Notice the analogy between quantum error correction schemes \cite{gottesman}, as depicted in \fig{fig:error_correction_scheme}, and the purification of measurements.
	For error correction, the message is first encoded by gate
        $\mathcal{R}$ into one of the carefully chosen codewords,
        which is then (possibly) corrupted by the noisy communication
        channel $\mathcal{E}$. Finally, in gate $\mathcal{D}$ some set of commuting Hermitian
        operators are measured over the corruption, the
        syndrome is used to perform error correction, and finally the recovered codeword is decoded
        into the original message.  For purification of measurements,
        we are allowed to encode the $N$ identical copies of input
        state $\rho_g$ through the channel $\mathcal{R}$, in a way
        similar to quantum error correction.  The aim of such encoding
        is very different, since after that we are forced to perform
        $M$ measurements with the same noisy POVM $\{P'_i\}$, which
        provide us just classical outcomes to be classically
        post-processed.  The limitation of the measurement purification
        versus error correction is that the decoding $\mathcal{D}$ is
        restricted to classical outcomes only.  The problem we are
        considering is also similar to the problem solved by
        entanglement purification protocols \cite{bennet}, since we
        are generally trying to recast the use of a number of noisy
        measurements to an effective use of a smaller number with less
        noise.
	
	The paper is organized as follows.  In
        \sect{sect:qubit_measurements} we specify the general problem
        to a qubit with isotropic noise, and then we face the
        optimization considering different figures of merit: in
        \sect{sect:variance_minimization} we show how to minimize the
        measurement noise, while in \sect{sect:mutual_information} we
        maximize the mutual information between the parameter
        describing the state and the outcomes of the POVMs.  In
        \sect{sect:photodetection} and
        \sect{sect:continuous_measurements}, we consider observables
        with unbounded spectrum, for which the concept of
        amplification applies, and we review the scheme of H. P. Yuen
        \cite{yuen} for purifying photodetectors
        (\sect{sect:photodetection}), homodyne and heterodyne
        detectors (\sect{sect:continuous_measurements}).  Finally,
        \sect{sect:conclusion} is devoted to conclusions.
	
	\section{Purification of Qubit Measurements}\label{sect:qubit_measurements}
	
\begin{figure}[htb]
		\includegraphics{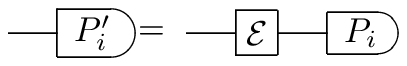}
		\caption{Noisy POVM element.}
		\label{fig:povm_element}
	\end{figure}

	Let us specify the general problem we are considering.  We are
        provided with $N$ identical copies of the input state $\rho_g$
        of dimension $d$. In what follows we will always suppose
        that the elements of the POVMs $\{P_i\}$ and
        $\{P'_i\}$ are $d$, which has been proved  to be the
        optimal choice for $d=2$ \cite{levitin}, when the mutual information is optimized \footnote{Indeed, for $d>2$ it has been shown by \cite{shor}, that a measurement with number of outcomes larger than the dimension of the span of the input states can improve the mutual information}.  We
        suppose that each noisy element $P'_i$ is obtained acting with
        the same channel $\mathcal{E}$ on the corresponding element
        $P_i$ of the ideal POVM
	\begin{equation}\label{eq:noisy_povm}
		P'_i = \mathcal{E}^\vee(P_i),
	\end{equation}
	where $\mathcal{E}^\vee$ denotes the Heisenberg-picture version of the channel $\mathcal{E}$.
        \eq{eq:noisy_povm} shows that the ideal POVM $\{P_i \}$ is ``cleaner'' that the noisy POVM
        $\{P'_i \}$ in the sense of the partial ordering introduced in \cite{buscemi05}, as depicted in
        \fig{fig:povm_element}.
	
	We consider a qubit (so $d=2$) parametrized as 
	\begin{equation}\label{eq:a_parametrization}
		\rho_{a,b} = \left(\begin{array}{cc}a & b\\b^* & 1-a\end{array}\right).
	\end{equation}
	We are interested in the observable $\sigma_z$, and we suppose to have at our disposal $M$ noisy POVM $\{P'_i\}$ of $\sigma_z$, i.e. $P_i = \ket{i}\bra{i}$. 
	We assume a simple kind of noise acting on each POVM, i.e. the
        isotropic noise
	\begin{equation}\label{eq:isotropic_noise}
		\mathcal{E}^\vee(P_i) = \alpha P_i + \beta I, 
	\end{equation}
	so that $P'_i = \alpha \ket{i}\bra{i} + \beta I$.
	
	We suppose to have $N=1$ qubit state and consider as a purification channel $\cal{R}$
        the orthogonal cloning $\mathcal{C}$, with respect to the basis of
        eigenstates of the observable $\sigma_z$
	\begin{equation}\label{eq:orthogonal_cloning}
		\mathcal{C} (\rho) = \sum_{i=0,1} \bra{i}\rho\ket{i} \ket{i}\bra{i}^{\otimes M}.
	\end{equation}
	
	The conditional probability $p(\vec{\imath}|a,b)$ of obtaining
        outcomes $\vec{\imath}=\{i_1,\dots,i_M\}$ given the state
        parametrized by $a$, $b$ does not depend on $b$, and can
        be explicitly written as
	\begin{equation}\label{eq:generic_conditional_probability}
		p(\vec{\imath}|a) = \Tr[\mathcal{C}(\rho)\mathcal{E}^\vee(P_i)^{\otimes M}]\;.
	\end{equation}
	
	We substitute \eq{eq:orthogonal_cloning} and \eq{eq:isotropic_noise} into \eq{eq:generic_conditional_probability} to obtain
	\begin{equation}
			p(\vec{\imath}|a) = \Tr[  (a \ket{0}\bra{0}^{\otimes M} + (1-a)\ket{1}\bra{1}^{\otimes M})
			 \otimes_{j=1}^M(\alpha\ket{i_j}\bra{i_j} +
                         \beta I) ]
	\end{equation}
        We observe that the probability $p(\vec{\imath}|a)$ depends only
        on the number of outcomes $0$'s and $1$'s in the measurement
        (i.e., not on their position).  Upon defining such integers as
        $M_0$ and $M_1=M-M_0$, we obtain
	\begin{equation}
			p(M_1|a) = {M \choose M_1} \left[ a(\alpha +
                          \beta)^{M_0}  \beta ^{M_1}
			 + (1-a)(\alpha + \beta)^{M_1}
                        \beta ^{M_0}\right] \;.
	\end{equation}
	
	For the normalization condition of the POVM in \eq{eq:isotropic_noise} one has $\alpha = 1 - 2\beta$, so $0 \le \beta \le \frac12$, and hence
	\begin{equation}\label{eq:conditional_probability}
			p(M_1|a) = {M \choose M_1} \left[ a \left((1-\beta)^{M_0} \beta^{M_1} - (1-\beta)^{M_1} \beta^{M_0}\right)
			+ (1-\beta)^{M_1} \beta^{M_0} \right ]\,.
	\end{equation}
	
	One can easily check the normalization of this probability,
        i.e. $\sum_{M_1=0}^M p(M_1|a) = 1$. In the case of ideal
        measurements for which $\beta =0$, the non-null probabilities
        are obtained just for $M_1=0$ and for $M_1=M$, namely
	\begin{equation}
		p(M_1=0|a) = a, \qquad p(M_1=M|a) = 1-a,
	\end{equation}
        whereas in the completely isotropic case (i.e.
        $\beta=\frac12$) the probability $p(M_1|a) = {M \choose M_1} \left(\frac12\right)^M$
	is independent of $a$, namely no information can be obtained
        about the state. 
	
	Notice that also the coherent channel, widely used in encoding schemes for quantum error correction as \cite{shor95},
	\begin{equation}
		\mathcal{C}' (\rho) = \sum_{i,j=0,1} \bra{i}\rho\ket{j} \ket{i}\bra{j}^{\otimes M}.
	\end{equation}
	leads to the same probability distribution
        \eq{eq:conditional_probability}, since $P'_i$ are diagonal on
        the $\sigma_z $ basis.
	
	\section{Minimization of Measurement Noise}\label{sect:variance_minimization}
	
	We show how to apply the ML criterion to obtain
        the optimal estimator for the expectation value of $\sigma_z$,
        by means of our measurement purification scheme.  Our aim is
        to show an improvement of estimation in terms of variance by increasing the uses of the POVM.
	In the following $\log$ will denote the logarithm to the base $2$.

	The ML criterion provides the following estimator for $a$ in the state \eq{eq:a_parametrization}
	\begin{equation}\label{eq:ml_estimator}
          a_{ML} = \arg \max_a \frac1n L(a|M_1)
	\end{equation}
	where $n$ is the number of (joint) outcomes (runs of the purification scheme depicted in \fig{fig:purification_scheme}), $L(a|M_1)$ is the so called log-likelihood functional
	\begin{equation}
		L(a|M_1) = \sum_{j=1}^n \log p_j(M_1|a),
	\end{equation}
	and $p_j(M_1|a)$ denotes the conditional probability for the $j$-th run.
	We observe that \eq{eq:ml_estimator} is concave since the
        logarithm of a linear function is a concave function and the
        summation of concave functions is a concave function.
	
	To solve the ML problem, we employ the iterative
        numerical method described in \cite{rao}.  First, we generate
        a large  amount of data distributed according to
        \eq{eq:conditional_probability}, for some fixed value of $a$
        and $\beta$.  Then, we fix some order zero approximation $a_0$
        for the estimator $a_{ML}$.  Then, the first order correction
        is given by
	\begin{equation}
		a_1 = \frac{\left.\frac{\partial L(a|M_1)}{\partial a}\right|_{a=a_0}}{F(a_0)}
	\end{equation}
	where $F(a)$ is the Fisher information
	\begin{equation}
		F(a) = \sum_{\vec{\imath}} \left(\frac{\partial{p(\vec{\imath}|a)}}{\partial{a}}\right)^2 \frac{1}{p(\vec{\imath}|a)},
	\end{equation}
	which in our case is given by 
	\begin{equation}
          F(a) = \sum_{M_1 = 0}^M {M \choose M_0} ^2\frac{\left((1-\beta)^{M_0} \beta^{M_1} - (1-\beta)^{M_1} \beta^{M_0}\right)^2}{p(M_1|a)}.
	\end{equation}
	The Fisher information measures the amount of information that
        the random variable $M_1$ carries about the unknown
        parameter $a$ on which the likelihood function depends.  So,
        the estimator to first order is $a_0 + a_1$, and the procedure
        can be iterated with this value as order zero approximation to
        obtain higher order corrections.  Obviously, the result is
        independent of the initial value $a_0$.  For $\beta$ not too big (say
        $0<\beta<\frac13$), the algorithm converges in a few steps
        (say, less than $10$).
	
\begin{figure}[htb]
 			\includegraphics[scale=.5]{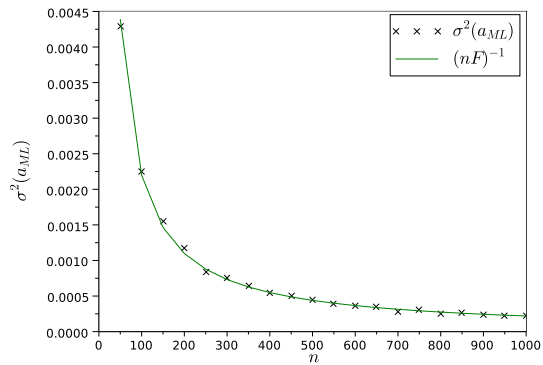}
 			\caption{(Color online) Variance $\var{a_{ML}}$ versus the
                          number $n$ of outcomes (runs), for parameter
                          $a=0.75$, measurement noise $\beta=0.25$,
                          and number of POVMs $M=10$. The solid line
                          represents the Cramer-Rao bound in
                          Eq. (\ref{eq:cramer_rao_bound}).}
 			\label{fig:cramer_rao_bound}
 	\end{figure}

	The variance on the ML estimator of a parameter satisfies the Cramer-Rao bound \cite{cramer}
	\begin{equation}\label{eq:cramer_rao_bound}
		\var{a_{ML}} \ge \frac{1}{n F(a)}
	\end{equation}
	The bound in \eq{eq:cramer_rao_bound} is saturated if the
        number of data is large enough and the parameter is
        mono-dimensional (as in the present case).  We numerically
        estimated the variance of the estimator $a_{ML}$ in \eq{eq:ml_estimator}
        by dividing the data into blocks, finding the
        estimator $a_i$ for each block, and then calculating the
        variance of such estimators, namely
	\begin{equation}
		\var{a_{ML}} = \sum_i (a_i - a_{ML})^2.
	\end{equation}
	In \fig{fig:cramer_rao_bound} we verified that the variance numerically saturates the bound in \eq{eq:cramer_rao_bound}.
	
\begin{figure}[htb]
		\includegraphics[scale=0.5]{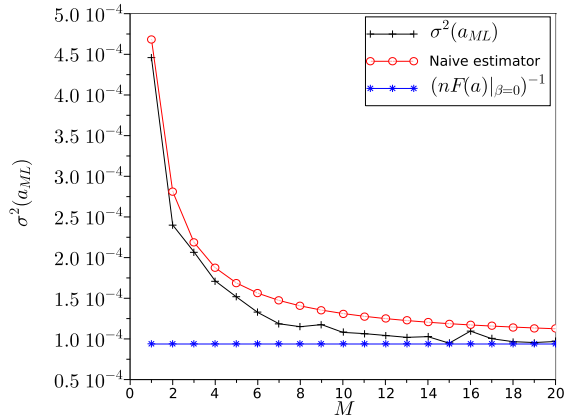}
		\caption{(Color online) Variance $\var{a_{ML}}$ versus the number $M$
                  of POVMs, for parameter $a=0.75$, measurement noise
                  $\beta=0.25$, and number of runs $n=2000$. The upper
                and lower bound correspond to Eqs. (\ref{eq:variance_upper_bound}) and
                (\ref{eq:variance_lower_bound}), respectively.}
		\label{fig:variance}
	\end{figure}

	\fig{fig:variance} shows that the variance decreases as the number of POVMs used in parallel increases, and upper and lower bounds for variance.
	To find the upper bound consider the function
	\begin{equation}
		f(M_1) = \frac{\frac{M_1}{M}+\beta -1}{2\beta-1}.
	\end{equation}
	Notice that $f(M_1)$ is an unbiased estimator for the parameter $a$, since one has
	\begin{equation}
		\expect{f(M_1)} = \sum_{M_1=0}^M f(M_1) p(M_1|a) = a.
	\end{equation}
	The second moment is given by
	\begin{equation}
		\begin{split}
			\expect{f(M_1)^2} & = \sum_{M_1=0}^M f(M_1)^2 p(M_1|a)\\
			& = a  +\frac{\beta (1-\beta ) }{(1-2\beta )^2 M}.
		\end{split}
	\end{equation}
	Thus, an upper bound for the variance on the parameter $a$ is
	\begin{equation}\label{eq:variance_upper_bound}
		\var{a_{ML}} \le \left(a-a^2  +\frac{\beta (1-\beta
                    )}{(1-2 \beta )^2 M}\right)\frac1n.
	\end{equation}
	The lower bound for the variance is
	\begin{equation}\label{eq:variance_lower_bound}
		\var{a_{ML}} \ge \frac{1}{n F(a)|_{\beta=0, M=1}}=
                \frac{a-a ^2}{n},
	\end{equation}
	where the right-hand side of \eq{eq:variance_lower_bound} corresponds to the use of the ideal detector on the original state.
	
	The computed variance shows a dependence on the
        parameter $a$ similar that in
        \eq{eq:variance_upper_bound}, decreasing as $1/M$ for $a=0$ or
        $a=1$.  The variance saturates the
        lower bound in \eq{eq:variance_lower_bound}, so the the
        estimator of the parameter $a$ is optimal.
	
	\section{Maximization of Mutual Information}\label{sect:mutual_information}
	
	We consider now the mutual information as the figure of merit
        in the measurement  purification scheme. 
	We consider a qubit parametrized as 
	\begin{equation}
		\ket{\psi} = \cos{\frac\theta2}\ket{0} + e^{i\phi}\sin{\frac\theta2}\ket{1}.
                \label{thph}
	\end{equation}

	The probability in \eq{eq:conditional_probability} can be written as
	\begin{equation}
			p(M_1|\theta) =  {M \choose M_1 }\left[
                        [(1-\beta)^{M_0} \beta^{M_1} - (1-\beta)^{M_1}
                        \beta^{M_0}] \cos ^2{\frac\theta2}  + (1-\beta)^{M_1} \beta^{M_0}\right],
	\end{equation}
	independent of $\phi$.
	
	In the following we suppose that the prior probability $p(\theta,
        \phi)$ of having the input state in Eq. (\ref{thph}) is uniform, so
        that the mutual information $I(M_1 : \theta, \phi)$ between random
        variables $\theta$ and $\phi$ and random variable $M_1$ is
        given by 
	\begin{equation}\label{eq:mutual_information}
					I(M_1 : \theta) :=  \frac12
                                        \int_0^\pi d\theta \sin\theta
                                        \sum_{M_1=0}^M p(M_1|\theta)
 \log\left(\frac{p(M_1|\theta)}{\frac12 \int_0^\pi d\theta \sin\theta p(M_1|\theta)}\right).
	\end{equation}
	The integral in the denominator gives 
	\begin{equation}
		\int_0^\pi d\theta \sin\theta p(M_1|\theta)= {M \choose
                  M_1}((1-\beta)^{M_0}\beta^{M_1} + (1-\beta)^{M_1}\beta^{M_0})\,,
	\end{equation}
	and a lengthy analytical form for \eq{eq:mutual_information}
        is provided in the Appendix of the paper. 
	
\begin{figure}[htb]
          \includegraphics[scale=.5]{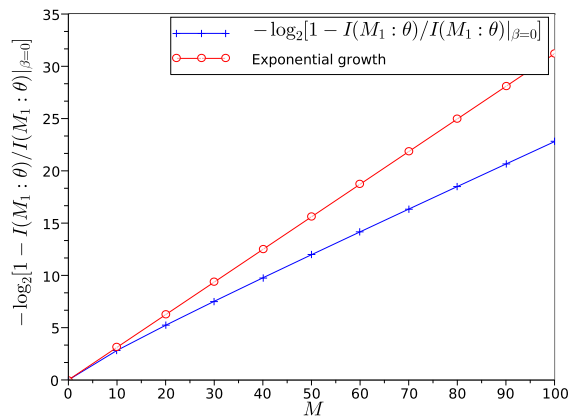}
          \caption{(Color online) Function
            $-\log _2\left
              [1-\frac{I(M_1:\theta)}{I(M_1:\theta)|_{\beta=0}}\right ]$
            versus the number $M$ of POVMs, for measurement noise
            $\beta=0.25$. The mutual information $I(M_1 : \theta )$ is
          given by Eq. (\ref{A1}).}
 		\label{fig:mutual_information}
 	\end{figure}

	The mutual information $I(M_1 : \theta)$ saturates the bound
        $I(M_1:\theta) \le I(M_1:\theta)|_{M=1,\beta=0} \simeq
        0.279\bit$ for increasing M.  Notice that the mutual information does not
        converge to $1\bit$, since a continuous ``alphabet'' of
        states is 
        allowed.  The mutual information $I(M_1:\theta)$ saturates almost exponentially versus $M$, as shown by \fig{fig:mutual_information}.
	This means that we are recasting the use of many noisy detectors to an effective use of a single ideal detector. 
	
\begin{figure}[htb]
		\includegraphics[scale=.5]{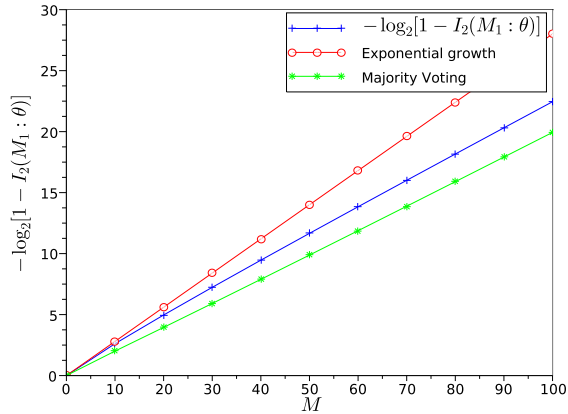}
		\caption{(Color online) Function $-\log _2[1-I_2(M_1:\theta)]$ versus the number $M$ of POVMs, for measurement noise $\beta=0.25$.}
		\label{fig:majority_voting}
	\end{figure}

	Let us consider in more detail the simplified case in which
        the only allowed input state are the up
        ($\theta=\pi$) and down ($\theta=0$) eigenstates of
        $\sigma_z$.  This simplification leads to two advantages: a
        much more tractable analytical form for the mutual information
        $I_2(M_1 : \theta)$, and the possibility to make a comparison
        with classical post-processing based on majority voting.  The mutual information
        is given by 
	\begin{equation}\label{eq:binary_mutual_information}
                  I_2(M_1 : \theta) = \sum_{M_1=0}^M  {M \choose M_1} (1-\beta)^{M_1}\beta^{M_0}
                  \log\left[\frac{2(1-\beta)^{M_1}\beta^{M_0}}{(1-\beta)^{M_0}\beta^{M_1}+(1-\beta)^{M_1}\beta^{M_0}}\right].
	\end{equation}
	\eq{eq:binary_mutual_information} behaves as expected for the ideal POVM case (i.e. $\beta=0$), where $I_2(M_1:\theta) = 1$, and for completely isotropic POVM case (i.e. $\beta=\frac12$) where $I_2(M_1:\theta) = 0$.
	Finally, we investigate the optimal classical
        post-processing to be applied on the $M$ outcomes of the
        parallel noisy POVMs to maximize the mutual information.  We
        simply argue that majority voting is close to the optimal
        post-processing, as is shown in plot \fig{fig:majority_voting}.
	The gap between the binary mutual information $I_2(M_1:\theta)$ and that obtained with majority-voting strategy could be explained
        by the fact that in general a number of
        POVM elements greater than the cardinality of the input alphabet can optimize the mutual information. In fact, Davies'
        theorem \cite{davies} puts an upper bound of $d^2$ on the number of
        POVM elements to optimize the mutual information for an alphabet of $d$ linear independent pure states (see also \cite{shor}).
	
\begin{figure}[htb]
		\includegraphics[scale=.8]{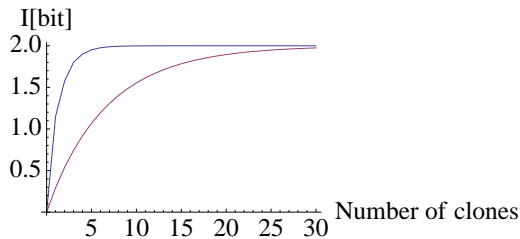}
		\caption{(Color online) Mutual information for an alphabet of four
                  equiprobable orthogonal states versus number of
                  purifying copies, for measurement noise $\alpha=0.8$
                  (upper), and $\alpha =0.4$ (lower).} 
		\label{fig:qudit4}
	\end{figure}

The case of two-states alphabet can be easily generalized to an
alphabet of $d$ orthogonal state $\{|j \rangle \}$, with
$j=1,2, \ldots d$, and noisy POVM elements $P'_i = \alpha |i\rangle \langle i |
+\frac{1-\alpha }{d}I$. The conditional probability of the outcomes of $M$
noisy measurements on $M$ copies of $|j \rangle $ is simply the
multinomial 
\begin{eqnarray}
p(M_1,M_2,\ldots , M_{d-1}|j )= \frac{M !}{M_1 ! M_2 ! \cdots M_d}
\frac{[(d-1)\alpha  +1]^{M_j} (1-\alpha )^{M-M_j}}{d^M}
\;,\label{condpj}
\end{eqnarray}
where $M_l$ is the number of outcomes $l$ in the string of $M$
outcomes, and $M_d =M -\sum_{j=1}^{d-1 }M_j$. The conditional
probability allows one to evaluate the mutual information, and for 
increasing number of clones $M$, the noisy measurements are
purified. In Fig. \ref{fig:qudit4} we show the purification effect for 
an alphabet of four orthogonal equiprobable states, and two different
values of noise. As expected, for increasing value of $M$ the mutual
information approaches two bits. 
\section{Inefficient photodetection}\label{sect:photodetection}
In the rest of the paper we consider observables with unbounded
spectrum, for which the concept of amplification applies, and we
review the scheme of H. P. Yuen \cite{yuen} for improving noisy 
photodetectors, homodyne and heterodyne detectors. In the original
proposal of Ref. \cite{yuen} the signal-to-noise ratio improvement was
studied for noisy measurements with preamplification assistance. By
reviewing the results here, we explicitly consider the effect of
amplification as a purification of the noisy POVMs, and hence of the
outcome probability distributions.   

Light is revealed by exploiting its interaction with
atoms/molecules or electrons in a solid, and, essentially, each photon
ionizes a single atom or promotes an electron to a conduction band,
and the resulting charge is then amplified to produce a measurable
pulse. In practice, however, available photodetectors are not
ideally counting all photons, and their performances is limited by a
non-unit quantum efficiency $\eta$, namely only a fraction $\eta$ of
the incoming photons lead to an electric signal, and ultimately to a
{\em count}: some photons are either reflected from the surface of the
detector, or are absorbed without being transformed into electric pulses.  
Let us consider a light beam entering a photodetector of quantum
efficiency $\eta$, {\em i.e.} a detector that transforms just a
fraction $\eta $ of the incoming light pulse into electric signal. 
We will focus our attention to the case of the radiation field 
excited in a stationary state of a single mode at frequency $\omega $. 
Then, the Poissonian process of counting gives the following
probability $p_\eta (m)$ of revealing $m$ photons \cite{kelley}
\begin{eqnarray}
p_\eta (m) = \hbox{Tr}\left[\rho \,
\mbox{{\bf :}}{\frac {(\eta a^{\dag} a)^m}{m!}}\exp
(-\eta a^{\dag} a)\mbox{{\bf :}} \right] 
\label{pc-singlemode}\;,
\end{eqnarray}
where $\rho $ represents the quantum state of light, and $\mbox{{\bf :}}\ \mbox{{\bf :}}$ 
denotes the normal ordering of field operators. 

Using the identities
\begin{eqnarray}
&&\mbox{{\bf :}}({a^\dag } a)^n \mbox{{\bf :}}
=({a^\dag } )^n a^n ={a^\dag } a({a^\dag } a-1)\ldots({a^\dag }
a-n+1)\;,\label{recurr}\\
&&
\mbox{{\bf :}}e^{-x{a^\dag } a}\mbox{{\bf :}}=
\sum_{l=0}^{\infty}{\frac{(-x)^l}{l!}}({a^\dag })^l a^l
=(1-x)^{{a^\dag } a},\label{Louisell}
\end{eqnarray}
one obtains 
\begin{eqnarray}
p_\eta (m) =\sum_{n=m}^{\infty} \rho_{nn}
\left(\begin{array}{c} n\\m\end{array}\right) \eta^m (1-\eta )^{n-m}
\label{conv_n}\;,
\end{eqnarray}
where 
\begin{eqnarray}
\rho_{nn}\equiv \langle n|\rho | n \rangle 
=p_{\eta =1}(n)
\;.\label{Pn}
\end{eqnarray}
Hence, for unit quantum efficiency a photodetector measures the photon
number distribution of the state, whereas for non unit quantum
efficiency the output distribution of counts is given by a Bernoulli
convolution of the ideal distribution.  

\par The outcome distribution in Eq. (\ref{conv_n}) can be
equivalently described
by means of a simple model in which the realistic photodetector is
replaced with an ideal photodetector preceded by a beam splitter of
transmissivity $\tau\equiv\eta$. The reflected mode is absorbed, whereas the
transmitted mode is photodetected with unit quantum efficiency.  In
order to obtain the probability of measuring $m$ clicks, notice that,
apart from trivial phase changes, a beam splitter of transmissivity
$\tau$ affects the unitary transformation of fields
\begin{eqnarray}
\left(\begin{array}{c} c\\d\end{array}\right) \equiv
U_\tau^\dag\left(\begin{array}{c} a\\b\end{array}\right)U_\tau= 
\left(\begin{array}{cc} \sqrt \tau& -\sqrt{1-\tau}\\
\sqrt{1-\tau}& \sqrt\tau \end{array}\right)
\left(\begin{array}{c}
a\\b\end{array}\right)\;,\label{bse}
\end{eqnarray} 
where all field modes are considered at the same frequency. 
Hence, the output mode $c$ hitting the detector is given by the 
linear combination
\begin{eqnarray}
c=\sqrt \tau  a-\sqrt{1-\tau} b\;,\label{a''}
\end{eqnarray}
and the probability of counts reads
\begin{eqnarray}
p_\tau (m) &=& \hbox{Tr} \left[ U_\tau   
(\rho\otimes|0\rangle \langle  0|)  
U^{\dag}_\tau   |m\rangle \langle m| \otimes 1 \right] \nonumber\\
&=&\sum_{n=m}^{\infty}\rho_{nn}
\left(\begin{array}{c} n\\m\end{array}\right) (1-\tau)^{n-m} \tau^m
\label{conv_n1}\;.
\end{eqnarray}
Equation (\ref{conv_n})  is then reproduced for $\tau=\eta$. We conclude that a photodetector
of quantum efficiency $\eta$ is equivalent to a perfect photodetector
preceded by a beam splitter of transmissivity $\eta$ 
which accounts for the overall losses of the detection process.
According to Eq. (\ref{conv_n}),  the POVM describing the inefficient
photodetector can be written as 
\begin{eqnarray}
\Pi _\eta (m) =
\left(\begin{array}{c} a^\dag a \\m \end{array}\right) \eta^m (1-\eta
)^{a^\dag a -m}
\;,\label{pietam}
\end{eqnarray}
such that $p_\eta (m)= \Tr [\rho \Pi _\eta (m)]$. 
The random variable $m$, suitably rescaled by $\eta $,  provides an estimator of the
average photon number $\langle a^\dag a \rangle =\Tr [\rho a^\dag a
]$, since one has 
\begin{eqnarray}
\sum _{m=0 }^ \infty \frac {m}{\eta } \Pi_\eta (m) = a^\dag a
\;. 
\end{eqnarray}
In order to evaluate the second moment of the probability, one uses
the identity
\begin{eqnarray}
\sum _{m=0 }^ \infty \left (\frac {m}{\eta }\right )^2 \Pi_\eta (m) = (a^\dag a)^2
+ \frac{1-\eta}{\eta }a^\dag a
\;, 
\end{eqnarray}
and hence the inefficient measurement is affected by the added noise
$\frac{1-\eta}{\eta }\langle a^\dag a \rangle $, with respect to the
ideal intrinsic noise $\Delta (a^\dag a)^2 \equiv \langle (a^\dag a)^2
\rangle - \langle a^\dag a\rangle ^2$. 
\par In the following we show that and ideal 
photon-number amplifier can arbitrarily reduce the added noise of the
inefficient measurement for increasing gain. 
The ideal photon-number amplification map is given by 
\cite{pna11,pna2,pna3} 
\begin{eqnarray}
a^{\dag }a \longrightarrow \hat V^{\dag }\,a^{\dag }a\,\hat
V=g\,a^{\dag }a\;,
\end{eqnarray}
where $g$ is an integer, and $\hat V$  is the isometry 
\begin{eqnarray}
\hat V=\sum_{n=0}^{\infty }|gn\rangle \langle n|\;. 
\end{eqnarray}
The preamplified POVM is simply given by 
\begin{eqnarray}
\Pi ^{(g)}_\eta (m) =
\left(\begin{array}{c} g a^\dag a \\m \end{array}\right) \eta^m (1-\eta
)^{g a^\dag a -m}\;.\label{pietamg}
\end{eqnarray}
The estimator of the average photon number is now $m/(g \eta )$, 
and the second moment is given by 
\begin{eqnarray}
\sum _{m=0 }^ \infty \left (\frac {m}{g\eta }\right )^2 \Pi ^{(g)}_\eta (m) = (a^\dag a)^2
+ \frac{1-\eta}{ g\eta }a^\dag a
\;. 
\end{eqnarray}
Clearly, for $g\to \infty $, the added noise is completely removed for
any value of the quantum efficiency $\eta $. 
\par We notice that the ideal photon-number amplifier is so effective
that indeed even a preamplified heterodyne detection provides the
ideal photon number distribution for increasing gain, as shown in
Ref. \cite{dys}. 

\section{Inefficient continuous variable measurements}\label{sect:continuous_measurements}
\subsection{Homodyne detection}\label{homsec}
The balanced homodyne detector provides the measurement of the
quadrature of the field 
\begin{eqnarray}
X_{\varphi }=\frac {{a^\dag } e^{i\varphi }+ a e^{-i \varphi }}{2}\;\label{xfi}
\end{eqnarray}
It was proposed by Yuen and Chan \cite{yuenchan}, 
and subsequently experimentally demonstrated by Abbas, Chan and Yee
\cite{abbas}. The signal mode $a$ interferes 
with a strong laser beam mode $b$ in a balanced 50/50 beam splitter. 
The mode $b$ is the so-called the {\em local oscillator} (LO) 
mode of the detector. It operates at the same frequency of $a$, 
and is excited by the laser in a strong coherent state $|z\rangle $.
Since in all experiments that use homodyne detectors the signal and the LO 
beams are generated by a common source, we assume that they have a fixed 
phase relation. In this case the LO phase provides a reference for the 
quadrature measurement, namely  we identify the 
phase of the LO with the phase difference between the two modes. 
By tuning $\varphi =\arg z$ we can measure the quadrature $X_\varphi $ at 
arbitrary phase. 

Behind the beam splitter, the two modes are detected by two identical
photodetectors (usually linear avalanche photodiodes), and finally the
difference of photocurrents at zero frequency is electronically
processed. In the strong-LO limit $|z|\to\infty$, the homodyne
detector is described by the POVM
\begin{eqnarray}
\Pi (x)= \int_{-\infty}^{+\infty} {\frac {d \lambda }{2\pi}} \,
\exp[i\lambda  (X_\varphi -x)]=
|x\rangle _\varphi {}_\varphi \langle x|\;,\label{pmx}
\end{eqnarray}
namely the projector on the eigenstate of the quadrature $X_\varphi  $
with eigenvalue $x$.  In conclusion, the balanced homodyne detector
achieves the ideal measurement of the quadrature $ X_\varphi $ in the
strong LO limit.  In this limit, the probability distribution of the
output photocurrent approaches exactly the probability
distribution $p(x,\varphi  )={}_\varphi \langle x|\rho |x \rangle  _\varphi  $ 
of the quadrature $X_ \varphi  $, and this for
any state $\rho $  of the signal mode $a$. 
\par It is easy to take into account non-unit quantum
efficiency at detectors. The POVM is obtained by replacing 
\begin{eqnarray}
X_\varphi \rightarrow X_\varphi +\sqrt{\frac{1-\eta }{2\eta
}}(u_\varphi +v_\varphi)\;
\end{eqnarray}
in Eq. (\ref{pmx}), with $w_\varphi =(w^\dag e^{i\varphi} +w
e^{-i\varphi})/2 $, where $w=u,v$ denotes the vacuum modes of the two
inefficient photodetectors. By tracing the vacuum modes
$u$ and $v$,  one obtains
\begin{eqnarray}
\Pi _{\eta}(x)&=&\int _{-\infty }^{+\infty}
\frac{d\lambda}{2\pi}\,e^{i\lambda  (X_\varphi -x)}
|\langle 0|e^{i\lambda
\sqrt{\frac{1-\eta}{2\eta}}u_{\varphi }}|0\rangle|^2 \\
&=&\int _{-\infty }^{+\infty}\frac{d\lambda}{2\pi}\,
e^{i\lambda (X_\varphi -x)}
e^{-\lambda^2\frac{1-\eta}{8\eta}}\nonumber\\
&=&\frac{1}{\sqrt{2\pi\Delta^2_{\eta}}}
\exp\left[ -\frac{(x-  X_\varphi )^2}{2\Delta_{\eta}^2}\right]
\nonumber \\&= &\frac{1}{\sqrt{2\pi\Delta^2_{\eta}}}
\int_{-\infty }^{+\infty}dx'\, e^{ -\frac{1}{2\Delta_{\eta}^2}
(x-  x')^2}\,|x'\rangle _\varphi {}_\varphi \langle x'| \nonumber 
\;,\label{OM}
\end{eqnarray}
where 
\begin{eqnarray}
\Delta_{\eta}^2=\frac{1-\eta}{4\eta}\;.\label{Deltaeta}
\end{eqnarray}
Thus the noisy POVM, and in turn the probability distribution of the output 
photocurrent, are just the Gaussian convolution of the ideal ones 
with r.m.s. $\Delta_{\eta}=\sqrt{(1-\eta )/(4\eta )}$.
\par In the following we show that the added noise of the inefficient
homodyne detector can be removed by amplifying the signal by means of
a phase-sensitive amplifier. This amplifier is described by the
squeezing operator 
\begin{eqnarray}
S(\xi )=\exp \left[ \frac 12 \left (\xi  a^{\dag 2} - \xi^* a^2
  \right) \right ]\;, 
\end{eqnarray}
and performs the mode transformation
\begin{eqnarray}
S^\dag (\xi ) a S(\xi )=(\cosh |\xi |) a + \frac{\xi }{|\xi |}(\sinh
|\xi |) a^\dag \;.
\end{eqnarray}
For $\xi = r  e^{2 i \varphi}$, with $r>0$, one has 
\begin{eqnarray}
S^\dag (\xi ) X_\varphi  S(\xi )= e^r X_\varphi \;.
\end{eqnarray}
Hence, the effective POVM obtained by preprocessing $\Pi _\eta (x)$ in
Eq. (\ref{OM}) with the phase-sensitive amplification of $X_\varphi $
is given by 
\begin{eqnarray}
\Pi ^{(r)}_\eta (x)  &=&S^\dag (r e^{2i\varphi }) \Pi _\eta (x) S(r e^{2i\varphi }) 
\nonumber \\&= &  \frac{1}{\sqrt{2\pi\Delta^2_{\eta}}}
\exp\left[ -\frac{(x-  e ^r X_\varphi )^2}{2\Delta_{\eta}^2}\right]
\;. 
\end{eqnarray}
Now, in order to obtain an unbiased measurement of $X_\varphi $, it
is enough to rescale the outcome by $e^{r}$. On the other hand, the
added noise with respect to the ideal measurement $X_\varphi$ becomes
equal to $e^{-2 r} \Delta ^2 _\eta $, which can be made arbitrary
small for increasing value of the squeezing parameter $r$. 

\subsection{Heterodyne detection}\label{hetsec} 
Heterodyne detection allows to perform the joint measurement of two
conjugated quadratures of the field \cite{hetyuen1,hetyuen2}. 
\par A strong local oscillator at frequency $\omega $ in a coherent
state $|\alpha \rangle $ hits a beam splitter with transmissivity
$\tau \to 1$, and with the coherent amplitude $\alpha $ such that $\gamma  
\equiv |\alpha |\sqrt {\tau (1-\tau)}$ is kept constant.  If the
output photocurrent is sampled at the intermediate frequency $\omega
_{IF}$, just the field modes $a$ and $b$ at frequency $\omega \pm
\omega _{IF}$ are selected by the detector.  Modes $a$ and $b$ are
usually referred to as signal band and image band modes,
respectively.  In the strong LO limit, upon tracing the LO mode, 
the output photocurrent $I(\omega _{IF})$ rescaled by $\gamma $ 
is equivalent to the complex operator 
\begin{eqnarray}
Z=\frac{I(\omega _{IF})}{\gamma }=a-b^\dag,\label{zeta}
\end{eqnarray}
 where the arbitrary phases of modes have been suitably chosen.
The heterodyne photocurrent $Z$ is a normal operator, 
equivalent to a couple of commuting
self-adjoint operators 
\begin{eqnarray}
Z=\hbox{Re}Z +i \hbox{Im}Z\;,\qquad [Z,Z^\dag ]= 
[\hbox{Re}Z,\hbox{Im}Z]=0\;.
\; 
\end{eqnarray}
The POVM of the detector is then given by the orthogonal (in Dirac
sense) eigenvectors of $Z$.
\par In conventional heterodyne detection the image band mode is in the
vacuum state, and one is just interested in measuring the field mode
$a$. In this case the POVM $\Pi (z )$ is obtained upon tracing on mode $b$, and  
one has the customary projectors on coherent states 
\begin{eqnarray}
\Pi (z)=
\frac{1}{\pi} |z \rangle \langle z| 
\;,   
\end{eqnarray}
with $z \in {\mathbb C}$. The coherent-state POVM provides the optimal joint measurement of
conjugated quadratures of the field \cite{hel}. For a state $\rho $, 
the expectation value of any quadrature $X_\varphi $ is obtained as
\begin{eqnarray}
\langle X_\varphi \rangle =\hbox{Tr}[\rho X_\varphi ]= \int _{\mathbb
C}\frac{d^2 \alpha}{\pi } \hbox{Re}(\alpha e^{-i \varphi}) 
\langle \alpha | \rho | \alpha \rangle   \;.
\end{eqnarray}
The price to pay for jointly measuring non-commuting observables is an
additional noise. The r.m.s. fluctuation is evaluated as follows
\begin{eqnarray}
\int_{\mathbb C} \frac{d^2 \alpha}{\pi } 
[\hbox{Re}(\alpha e^{-i \varphi})]^2  \langle \alpha | \rho | \alpha \rangle   
- \langle X_\varphi \rangle ^2 =
\langle \Delta X^2 _\varphi \rangle +\frac {1}{4}
\;, 
\end{eqnarray}
where $\langle \Delta X^2 _\varphi \rangle $ is the intrinsic noise,
and the additional term is usually referred to as ``the additional 
3dB noise due to the joint measure'' \cite{Arthurs,yuen82,goodman}.
\par The effect of non-unit quantum efficiency can be taken into account
in analogous way as in Sec. \ref{homsec} for homodyne detection. 
The coherent-state POVM is replaced with the convolution
\begin{eqnarray}
\Pi _\eta (z)= 
\int _{\mathbb C}\frac{d^2 z'}
{\pi \Delta ^2_\eta }e^{-\frac {|z'-z|^2}{\Delta ^2_\eta } }
\frac{|z' \rangle \langle z'|}{\pi}\;,
\label{coeta}
\end{eqnarray}
where $\Delta^2_{\eta}=(1-\eta )/\eta $. 
\par In the following we show that inefficient heterodyne detection can be
purified by  phase-insensitive amplification. 
Phase-insensitive amplification with (power) gain G amplifies the
coherent amplitude of coherent states by $\sqrt G$, at the expense of
addition thermal photons $\bar n = G-1$. Differently from
phase-insensitive amplification, the physical process is not unitary,
but described by a completely positive map ${\cal E} _G$. Here, we just need the
Heisenberg evolution of the projector on coherent states, which is
simply given by the rescaling \cite{caves}
\begin{eqnarray}
{\cal E}_G ^{\vee } (|\alpha \rangle\langle \alpha |)=
\frac 1G  |G^{-1/2}\alpha \rangle \langle G^{-1/2} \alpha | 
\;.
\end{eqnarray}
It follows that under phase-insensitive preamplification the noisy
heterodyne POVM (\ref{coeta}) is replaced with 
\begin{eqnarray}
\Pi ^{(G)}_\eta (z)
= \int _{\mathbb C}\frac{d^2 z'}
{\pi \Delta ^2_\eta }e^{-\frac {|\sqrt G z'-z|^2}{\Delta ^2_\eta } }
\frac{|z' \rangle \langle z'|}{\pi}\;.
\label{coeta2}
\end{eqnarray} 
Upon rescaling $z \to G^{-1/2}z$, one obtains 
\begin{eqnarray}
\Pi ^{(G)}_\eta (z)
&& =  \int _{\mathbb C}\frac{G d^2 z'}
{\pi \Delta ^2_\eta }e^{-\frac {G| z'-z|^2}{\Delta ^2_\eta } }
\frac{|z' \rangle \langle z'|}{\pi}\;\nonumber \\&& = _{\scriptsize{G\to \infty}}  \frac{|z \rangle \langle z|}{\pi}
\;,
\end{eqnarray}
namely the noise due to quantum efficiency can be arbitrarily reduced
for increasing value of the gain $G$. 
The effectiveness of preamplification in purifying heterodyne detection is
more unexpected than the case of homodyne detection, since
phase-insensitive amplification is not a unitary process.
	
	\section{Conclusion}\label{sect:conclusion}
	
	In this paper, we addressed the problem of optimizing a
        quantum channel acting before many parallel uses of a noisy
        POVM in order to purify the measurements, namely to achieve an
        effective measurement that is less noisy than the original
        ones. We first considered
		the purification of $\sigma_z$ noisy measurements on qubits, by choosing the orthogonal cloning channel as a purification map.
        We found the maximum-likelihood estimator for $\sigma_z$, whose variance shows an
        almost-exponential decay of versus the number of
        POVMs.  We also worked out an analytic form for the mutual
        information between the state parameter and the outcomes of
        the POVMs, and here also an almost-exponential improvement
        versus the number of POVMs has been found.  We proved that
        naive majority voting is not the optimal classical
        post-processing, since the  maximum-likelihood approach
        gives a better estimator. For photodetection and continuous
        variable measurements as homodyne and heterodyne detection,
        the measurement purification can be achieved by
        preamplification, as early pointed out by H. P. Yuen
        \cite{yuen}. 
\par We think that the relevant problem of purifying
        noisy quantum measurements will have a significant impact on
        the quantum information technology, in this same way
        as the decoherence problem.

\section*{Acknowlegments}This work is supported by Italian Ministry of
Education through PRIN 2008 and the European Community through COQUIT project.

	\appendix\section{Derivation of the mutual information in
          Eq. (\ref{eq:mutual_information}) }\label{sect:derivation_mutual_information}
	
	We provide here an analytic form for
        \eq{eq:mutual_information}, obtained substituting
        \eq{eq:conditional_probability} for $p(M_1|\theta)$. One obtains 
	\begin{equation}
			I(M_1:\theta) = \frac1{32} \sum_{M_1=0}^M {M
                          \choose M_1} \frac{c_1 A + c_2 B}{C}, \label{A1}
	\end{equation}
	where
	\begin{equation}
		c_1 = (1-\beta)^{2M} \beta^{4M_1}, \qquad c_2 = (1-\beta)^{4M_1} \beta^{2M},
	\end{equation}
	and
	\begin{equation}
		\begin{split}
			A = & -16\log((1-\beta)^{2M_1} \beta^M + (1-\beta)^M \beta^{2M_1})\\
			& +8\log((1-\beta)^{2M} \beta^{4M_1}) +16 -8\frac1{\ln(2)},
		\end{split}
	\end{equation}
	\begin{equation}
		\begin{split}
			B = & +16\log((1-\beta)^{2M_1}\beta^M+(1-\beta)^M\beta^{2M_1})\\
			& -8\log((1-\beta)^{4M_1}\beta^{2M}) -16 +8\frac1{\ln(2)},
		\end{split}
	\end{equation}
	\begin{equation}
		C = (1-\beta)^{M+M_1}\beta^{3M_1}-(1-\beta)^{3M_1}\beta^{M+M_1}.
	\end{equation}
	
\end{document}